\documentclass[reprint,amsmath,amssymb,aps,prb]{revtex4-2}
\usepackage{graphicx}
\usepackage{dcolumn}
\usepackage{mathptmx}
\usepackage{bm}
\begin{document}
\title{Engineering Ferromagnetism in Wide Bandgap w-AlN for Spintronic Applications: Insights from DFT Calculations}
\author{Chinnappan Ravi}
\email{ravic@igcar.gov.in}
\affiliation{Materials Modelling Section, Defects and Damage Studies Division, Materials Science Group, Indira Gandhi Centre for Atomic Research, HBNI, 
Kalpakkam 603102, Tamil Nadu, India}
\date{\today}
\begin{abstract}
Spintronics drives the search for functional materials, pursuing room-temperature dilute magnetic semiconductors to realize practical, energy-efficient devices.
This paper presents a theoretical investigation into the magnetic properties of w-AlN doped with Cr, Ru, and Rh atoms, using spin-polarized density functional theory 
calculations with supercell models.  Calculations of point-defect formation energies as a function of the Fermi level predict that Cr$^{4+}$, Ru$^{4+}$, and Rh$^{3+}$ 
are the most favorable charge states substituting Al in w-AlN.  With these preferred charge states, Cr-doped AlN is stable in the ferromagnetic state, preferable for 
spintronic devices, across a wide concentration range (1.85 to 16.67\% of Al).  Conversely, Ru- and Rh-doped AlN are unstable in the ferromagnetic state relative to the 
antiferromagnetic state.  Further investigation into the electronic density of states reveals a fascinating evolution: for Cr concentrations below 5.56\%, the Fermi 
level resides within the band gap directly above the valence band maximum, keeping the system insulating.  It transitions to a high-spin half-metallic state between 
7.40\% and 12.96\%, and finally transforms into a normal metal at 16.67\% Cr.  This behavior is absent in the DOS of the antiferromagnetic model of Ru- and Rh-doped 
systems, where insulating and metallic behaviors instead appear non-sequentially with varying concentrations.
\end{abstract}
\maketitle

\section{\label{intro}INTRODUCTION}
Electrons can transport spin angular momentum through materials independently of their charge, giving rise to the physical phenomenon of spin 
current\cite{kukreja2015}.  Spintronics research focuses on the generation, manipulation, and detection of spin currents and spin-polarized electrons 
in insulators, semiconductors, and metals.  Spintronics extends conventional electronics, enabling novel devices with higher processing speeds, 
non-volatile data storage, lower power dissipation, and multifunctional capabilities\cite{fert2008,sato2010,hirohata2020,tanaka2021}.  A standard example 
of a spintronic device is magnetic random-access memory (MRAM), which utilizes multi-layer magnetic pillars to form standard giant magnetoresistance (GMR) 
stacks.  Within these pillars, interlayer magnetic coupling stores binary data that is subsequently read via the magnetoresistive response\cite{sato2010}.  

Another example of a spintronic device is a spin-polarized light-emitting diode (spin-LED), which utilizes a dilute magnetic semiconductor (DMS). 
In this device, Mn-doped GaAs produces holes that couple their spins to the magnetic moments of the Mn atoms, generating spin-polarized carriers. 
An n-type-doped region (InAs:Ge) is separated from this p-type region by a GaAs spacer layer.  Under an applied voltage, the spin-polarized holes 
migrate across the GaAs layer and recombine with electrons in the n-type region.  Due to the spin polarization of these holes, the resulting 
electroluminescence exhibits a specific helicity to satisfy angular momentum conservation.  Thus, the emission of circularly polarized light directly 
confirms the spin polarization of holes within the Mn-doped GaAs region\cite{sato2010}.  While this device currently operates only at low temperatures 
(170 K) and is not available commercially, spin-polarized LEDs offer potential for optical communications, quantum information processing, 
and biological imaging.

Ferromagnetic semiconductors could seamlessly integrate electronics with spintronics.  Optical and electrical manipulation of carrier-mediated 
ferromagnetism is possible in semiconductors due to their low carrier density, unlike in ferromagnetic metals.  A robust material exhibiting this unique 
functionality at room temperature and compatible with large‑scale manufacturing has not yet been discovered or synthesized.  The search for such materials, 
particularly in wide‑bandgap semiconductors, remains an active research field\cite{fert2008,sato2010,hirohata2020,tanaka2021}.  A popular technique for 
developing magnetic semiconductors is doping the host semiconductor with transition metals\cite{sato2010}.  The transition metal dopants introduce localized 
d states in the band gap.  These d electron spins tend to align forming stable magnetic moments that can be exploited in applications.  GaN:Mn was proposed 
as a promising candidate for room temperature ferromagnetism\cite{dietl2010}.  However, a recent study found no indication of high‑temperature ferromagnetic 
order in this material\cite{bonanni2011}.

Aluminum Nitride with the wurtzite structure (w-AlN) is thermodynamically more stable than GaN (formation enthalpy: -319.8 kJ/mol vs. -156.8 
kJ/mol)\cite{ranade2001}.  It is a direct wide-band gap semiconductor (6.2 eV; GaN: 3.4 eV)  with excellent thermal conductivity (321 W/mK), 
a high piezoelectric coefficient (5.1$\pm$0.1 pm/V), and superior acoustic velocity (6000 m/s)\cite{strite1992,cheng2020,lueng2000,terai2023}.  
AlN is utilized in optoelectronics such as lasers and ultraviolet light-emitting diodes\cite{frazier2005,wang2022,alan2023}, as well as in 
metal-insulator-semiconductor heterostructures (for power electronics and gas sensors) and in thermoelectric and piezoelectric 
devices\cite{frazier2005,wang2022,alan2023,zhou2017,startt2023,signore2024}.  Substantial p-type conductivity in AlN is also achieved by beryllium 
doping through metal-modulated epitaxy (MME)\cite{ahmad2021}; moreover, AlN with Cr substituted for Al is a promising candidate for semiconductor 
spin qubits\cite{czelej2024,burkard2023}.  Applications such as grid-level switching at tens of kV, ultraviolet emitters approaching the virus/bacteria 
extinction maximum of 204 nm, and high-temperature devices capable of operating without extensive thermal management require new semiconductors beyond 
GaN and SiC.  AlN and related alloys—with extreme bandgap, optical activity, bipolar doping, wafer sizes greater than 2 inch but scalable beyond 6 inch, 
and excellent thermal conductivity—provide the technological foundation necessary to drive widespread commercial adoption\cite{alan2023}.  These 
prospects make AlN a compelling material for dilute magnetic semiconductor spintronics as well.

Although theoretical and experimental studies of the magnetic properties of transition‑metal‑doped AlN exist, they are still incomplete.  
In particular, though several studies report room-temperature ferromagnetism in Cr-doped AlN, the variation of saturation magnetization 
with Cr concentration differs among studies.  Computational work often did not consider adequate configurations of dopants or their oxidation 
states.  Besides the 3d elements, 4d transition metals such as Ru and Rh have also been investigated as magnetic dopants in other nonmagnetic 
materials.  Monolayers of Ru and Rh interleaved in Ag, overlayers on Ag substrate, and clusters on the Ag surface have all been reported to be 
ferromagnetic\cite{blugel1992,wu1992,wildberger1995}.  Ma et al. have reported that the magnetic moments of most of the 3d and 4d dopants in silicon 
quantum dots are completely quenched, while the remaining spin moments on V, Cr, Mn, Nb, Mo, and Tc dopants are reduced in magnitude from those of 
respective free atoms\cite{ma2007}.  A study on the adsorption of transition metal atoms on single layer boron nitride shows that the transition metals 
(including Cr, Ru and Rh) exhibit nonzero magnetic moments\cite{li2018}.  Moreover, Ru is reported to be ferromagnetic at room temperature\cite{quarterman2018}.  
We have investigated the magnetic properties of Cr-, Ru-, and Rh-doped w-AlN to provide a comprehensive assessment of their magnetic order and 
electronic structure, which were not adequately addressed in our preliminary overview\cite{ravi2025}.  The results of this computational study verify that 
Cr-doped w-AlN is a promising candidate for dilute ferromagnetic semiconductors.  However, our findings do not favor Ru and Rh dopants for the development 
of w-AlN-based dilute magnetic semiconductor systems.  The paper is organized as follows.  Section~\ref{method} gives the details of the computational 
method briefly.  Results and discussions are presented in section~\ref{results}.  Conclusions are summarized in section~\ref{con}.

\section{\label{method}COMPUTATIONAL METHOD}
The thermodynamic properties of a point defect ($X$) are determined by its formation energy, $E^f(X^q)$.  The formation energy 
quantifies the energetic cost of introducing a defect in charge state $q$ under specific synthesis conditions.  The defect formation energy 
is expressed by\cite{prashun2025}:
\begin{equation}\label{eq1}
E^{f}(X^q)=E(X^q)-E(H)-\sum_{i}{n_i\mu_i}+qE_F+E_{corr}
\end{equation}
where $E(X^q)$ is the total energy of the supercell model of the semiconductor with defect $X$ in charge state $q$, and $E(H)$ is the total 
energy of the pure host supercell.  In this work, a $\mathrm{3\times3\times3}$ supercell of w-AlN consisting of 108 atoms (Al$_{54}$N$_{54}$) 
is used.  Here, $\mathrm{n_i}$ represents the number of atoms removed (+) from or added (-) to the host supercell to create the defect.
The chemical potential $\mu_i$ depends on the material's synthesis conditions and satisfies the boundary conditions:
$\mu_\mathrm{N}\leq\mu_{\mathrm{N_2}}$, $\mu_{\mathrm{Al}}\leq\mu_{\mathrm{Al}}^{\mathrm{bulk}}$, $\mu_{\mathrm{Cr}}\leq\mu_{\mathrm{Cr}}^{\mathrm{bulk}}$, 
$\mu_{\mathrm{Ru}}\leq\mu_{\mathrm{Ru}} ^{\mathrm{bulk}}$, and $\mu_{\mathrm{Rh}}\leq\mu_{\mathrm{Rh}}^{\mathrm{bulk}}$.  
We assume that the dopants and host elements are in thermal equilibrium with AlN, CrN, RuN, and RhN.  Under Nitrogen-rich synthesis condition\cite{fan2009}, 
the chemical potential of Al is determined by $\mu_{\mathrm{Al}}=E_{\mathrm{AlN}}^{\mathrm{bulk}}-\frac{1}{2}E_{\mathrm{N_2}}$.  The chemical 
potentials $\mu_{\mathrm{Cr}}$, $\mu_{\mathrm{Ru}}$, and $\mu_{\mathrm{Rh}}$ are obtained similarly, using the total energy of CrN, RuN, and RhN 
modeled in the rock-salt structure.  Finally, $E_F$ is the Fermi level (the chemical potential of electrons), and $E_{corr}$ is the finite-size 
correction for charged defects.  

The DFT total energy calculations for the charged defect formation energies were performed in two steps.  First, all structural degrees of freedom of the pure 
supercell were relaxed fully using the conjugate gradient algorithm.  Second, point defects (vacancy and dopant atoms) were introduced into the optimized structure, 
and the input files required to simulate the desired charge states were generated using the Python toolkit $\mathbf{doped}$\cite{kavanagh2024}.  The total energies 
of these defective systems were then calculated with fixed volume and lattice vectors, allowing only the ionic positions to relax.  The resulting energies were 
processed within $\mathbf{doped}$ to analyze the thermodynamic properties of the defects.  For the finite-size charge correction ($E_{corr}$), the anisotropic 
extended Freysoldt-Neugebauer-Van de Walle (eFNV) image-charge correction scheme was applied\cite{kumagai2014}, utilizing the mactroscopic dielectric constants 
of pure w-AlN obtained from the Materials Project database\cite{mp661}.

The DFT total energy calculations were carried out using the Vienna Ab initio Simulation Package\cite{kresse1993,kresse1996}.  Electron-ion interactions were 
described via the projector-augmented wave method\cite{kresse1999}, while the exchange-correlation functional was treated using the Perdew-Burke-Ernzerhof (PBE) 
parametrization of the generalized gradient approximation (GGA)\cite{perdew1996}.  Electron wave functions were expanded in a plane-wave basis set with a kinetic 
energy cutoff of 500 eV.  For Brillouin-zone sampling, a 7$\times$7$\times$5 Monkhorst-Pack k-point mesh was employed for the supercell models.  These specifications 
ensured that the total energies converged to better than 1 meV. 

In order to explore the magnetic properties of w-AlN doped with Cr, Ru and Rh atoms, spin polarized electronic band structure calculations were performed on supercell 
models of doped w-AlN, namely, Al$_{54-x}$M$_x$N$_{54}$, having dopants M = Cr, Ru or Rh in seven different values of $x$, namely, Al$_{53}$M$_1$N$_{54}$, 
Al$_{52}$M$_2$N$_{54}$, Al$_{51}$M$_3$N$_{54}$, Al$_{50}$M$_4$N$_{54}$, Al$_{49}$M$_5$N$_{54}$, Al$_{47}$M$_7$N$_{54}$, and Al$_{45}$M$_9$N$_{54}$.  These 
values of $x$ amounts to 1.85\%, 3.70\%, 5.56\%, 7.41\%, 9.26\%, 12.96\% and 16.67\% of Al.  This is the typical range of concentrations of transition metal 
dopants (M) examined in the computational study of DMS\cite{sato2010}.  Within these seven concentrations, 20 different configurations (patterns) of dopants 
(M) were considered.  We consider mainly nearest and next nearest neighbor configurations of dopants on the Al sublattice because the exchange interaction 
between dopants in DMS are generally short ranged\cite{sato2010,bonanni2011},  These configurations are described next.

\begin{figure}
\includegraphics[width=0.23\textwidth]{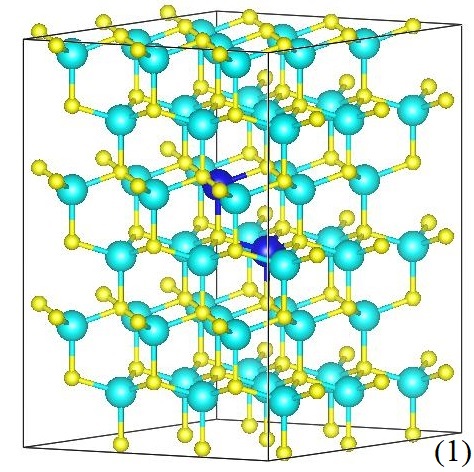}
\includegraphics[width=0.23\textwidth]{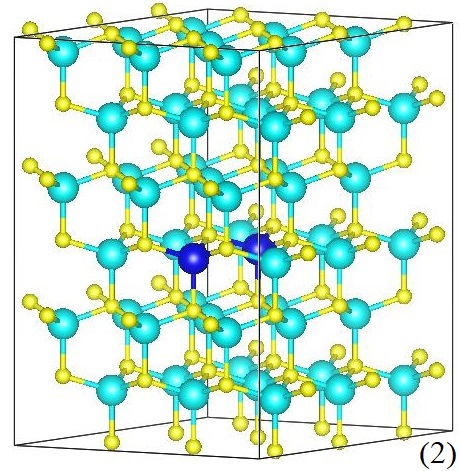}
\includegraphics[width=0.23\textwidth]{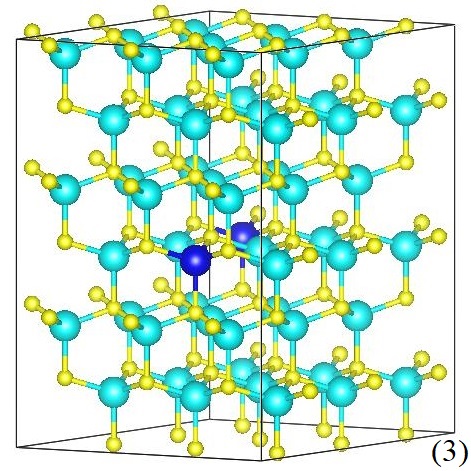}
\includegraphics[width=0.23\textwidth]{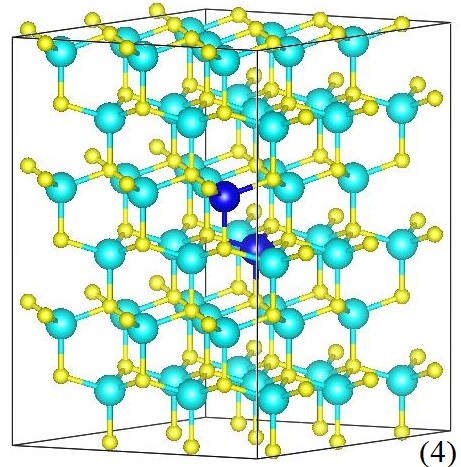}
\caption{\label{fig1}$\mathrm{3\times3\times3}$ supercell models of doped w-AlN: Al$_{52}$M$_2$N$_{54}$, where dopant atoms (M = Cr, Ru and Rh) substitute
Al sites.  Four distinct configurations were considered for the two M atoms.  Host Al and N atoms are represented in aqua and yellow, respectively, while the 
dopant atoms (M) are shown in blue.}
\end{figure}

Specifically, a single configuration was evaluated for an isolated dopant, while four, eight, three, and two configurations were studied for clusters of two, 
three, four, and five dopant atoms, respectively. For larger cluster sizes of seven and nine dopant atoms, one configuration was investigated in each case.
Fig.\ref{fig1}(1) depicts a configuration of two dopant atoms substituted for two nearest neighbor (nn) Al atoms distributed over two different $ab$ planes 
(bond length 3.08 $\text{\AA{}}$; planes of horizontal axes normal to the vertical axis).  Fig.\ref{fig1}(2) depicts a configuration of two nn dopant atoms 
distributed on a given $ab$ plane (3.12 $\text{\AA{}}$).  Figs.\ref{fig1}(3) and (4) show two next nearest neighbor (nnn) dopant atoms distributed 
respectively on a given $ab$ plane (5.40 $\text{\AA{}}$) and on two different $ab$ planes (4.38 $\text{\AA{}}$).

\begin{figure*}
\includegraphics[width=0.23\textwidth]{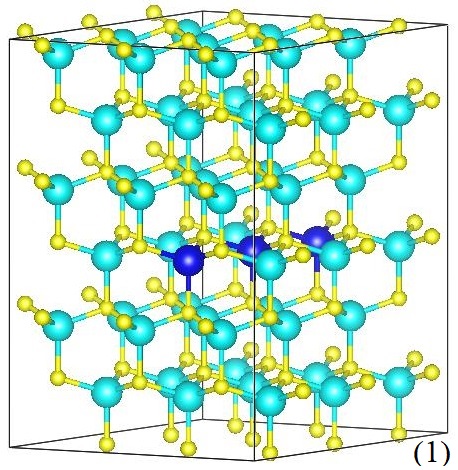}
\includegraphics[width=0.23\textwidth]{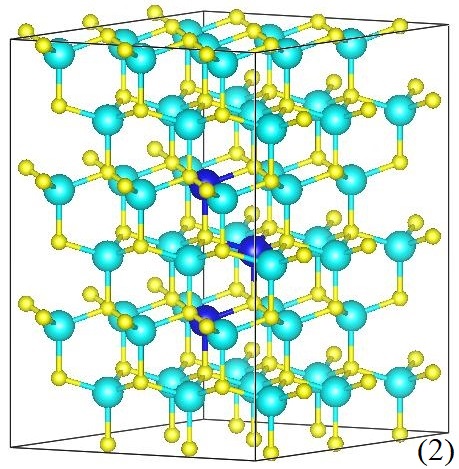}
\includegraphics[width=0.23\textwidth]{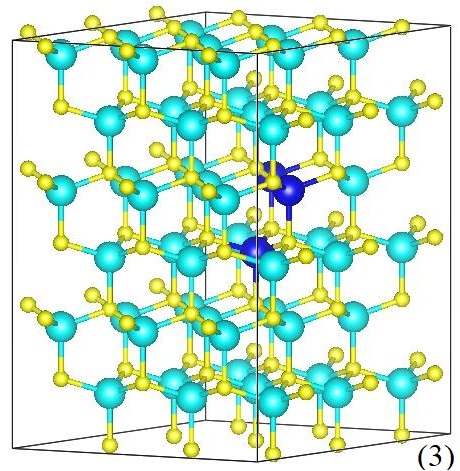}
\includegraphics[width=0.23\textwidth]{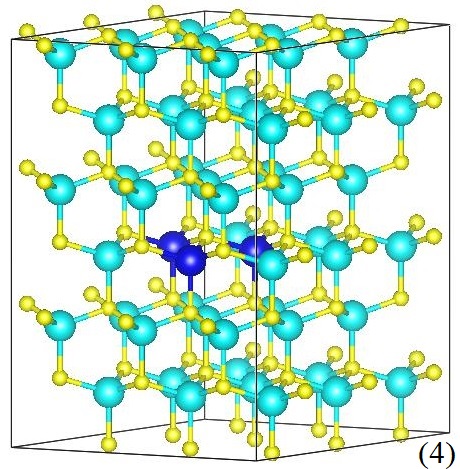}
\includegraphics[width=0.23\textwidth]{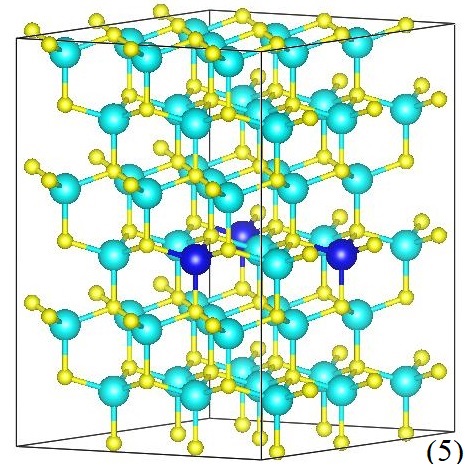}
\includegraphics[width=0.23\textwidth]{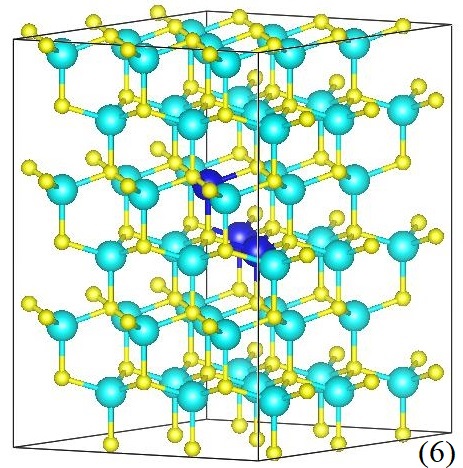}
\includegraphics[width=0.23\textwidth]{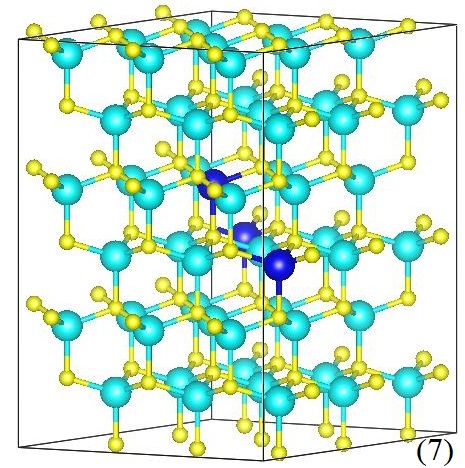}
\includegraphics[width=0.23\textwidth]{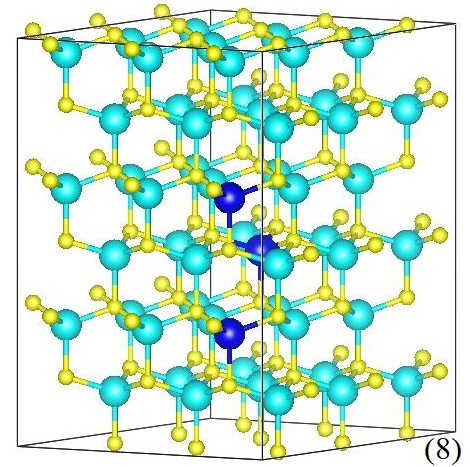}
\caption{\label{fig2}$\mathrm{3\times3\times3}$ supercell models of doped w-AlN: Al$_{51}$M$_3$N$_{54}$, where dopant atoms (M = Cr, Ru and Rh) substitute
Al sites.  Eight distinct configurations were considered for the three M atoms.  Host Al and N atoms are represented in aqua and yellow, respectively, while the 
dopant atoms (M) are shown in blue.}
\end{figure*}

Figs.\ref{fig2}(1) to (8) depict configurations of three dopant atoms. In panel (1), three nn collinear dopant atoms sit on a given $ab$ plane.  
Next, panel (2) represents three nn dopant atoms distributed on three different $ab$ planes, while panel (3) shows three nn dopant atoms distributed 
on two different $ab$ planes.  Panel (4) depicts a triangular configuration of three nn dopant atoms on a given $ab$ plane.  Moving to next-nearest neighbors, 
panel (5) depicts three nnn dopant atoms on a given $ab$ plane.  Panels (6) and (7) show three nnn dopant atoms distributed on two different $ab$ planes; 
however, panel (6) features three different bond lengths between pairs, whereas panel (7) features two.  Finally, panel (8) represents three nnn dopant 
atoms distributed over three $ab$ planes.

\begin{figure*}
\includegraphics[width=0.23\textwidth]{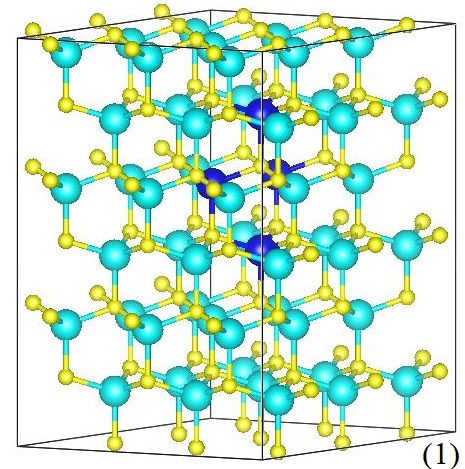}
\includegraphics[width=0.23\textwidth]{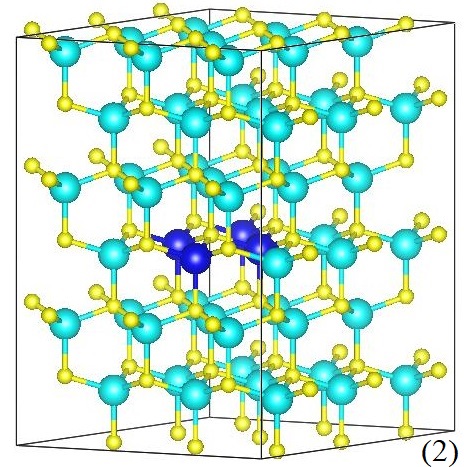}
\includegraphics[width=0.23\textwidth]{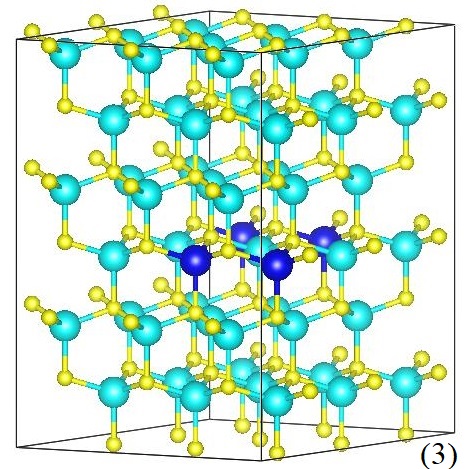}
\caption{\label{fig3}$\mathrm{3\times3\times3}$ supercell models of doped w-AlN: Al$_{50}$M$_4$N$_{54}$, where dopant atoms (M = Cr, Ru and Rh) substitute
Al sites.  Three distinct configurations were considered for the four M atoms.  Host Al and N atoms are represented in aqua and yellow, respectively, while the 
dopant atoms (M) are shown in blue.}
\end{figure*}

\begin{figure*}
\includegraphics[width=0.23\textwidth]{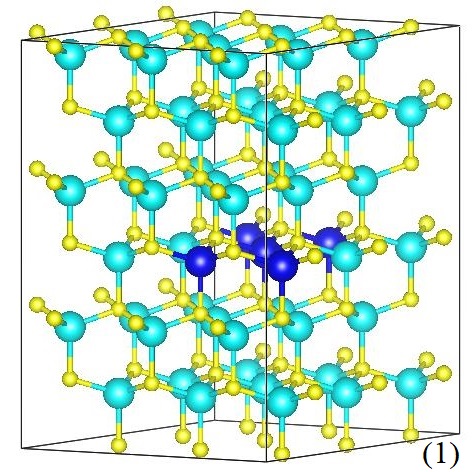}
\includegraphics[width=0.23\textwidth]{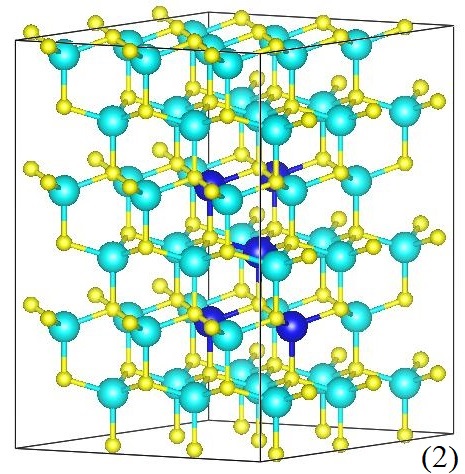}
\includegraphics[width=0.23\textwidth]{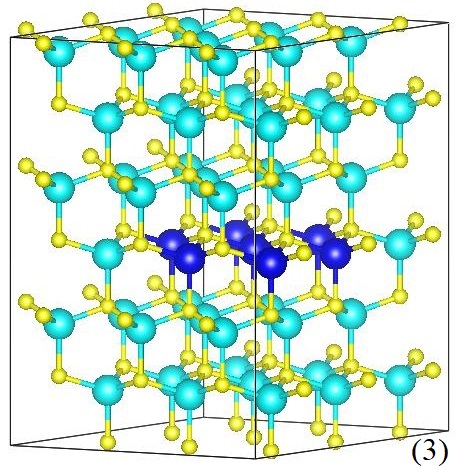}
\includegraphics[width=0.23\textwidth]{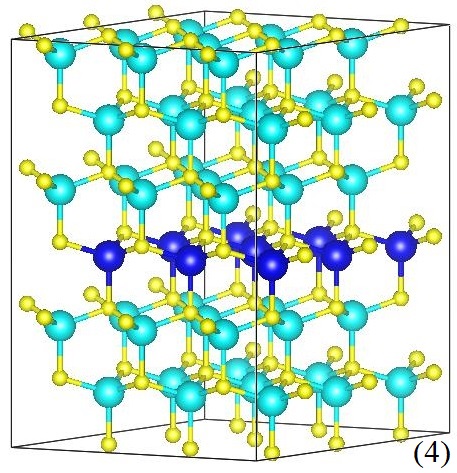}
\caption{\label{fig4}$\mathrm{3\times3\times3}$ supercell models of doped w-AlN: Al$_{49}$M$_5$N$_{54}$, Al$_{47}$M$_7$N$_{54}$, and Al$_{45}$M$_9$N$_{54}$, 
where dopant atoms (M = Cr, Ru and Rh) substitute Al sites.  Two distinct configurations were considered for the five M atoms.  With seven and nine M atoms, 
one configuration each was considered.  Host Al and N atoms are represented in aqua and yellow, respectively, while the dopant atoms (M) are shown in blue.}
\end{figure*}

Figs.\ref{fig3}(1) to (3) depict configurations of four dopant atoms.  Panels (1) and (2) represents four nn dopants distributed over three and one 
$ab$ planes respectively.  Panel (3)  depicts four nnn dopant atoms distributed on a given $ab$ plane.  Figs.\ref{fig4}(1) and (2) depict five dopant atoms 
distributed on one and three $ab$ planes respectively.  Figs.\ref{fig4}(3) and (4) show respectively seven and nine nn dopant atoms on a given $ab$ plane.  
The supercell model with a single dopant atom is not displayed here.  These structural details of dopant configurations will be useful for atom-by-atom 
fabrication prospects\cite{ding2023,yu2023}.

To determine the stable magnetic ground state of transition metal doped semiconductors, the total energy difference between the ferromagnetic (FM) and 
antiferromagnetic (AFM) states is calculated\cite{lany2008}.  Therefore, conventional ferromagnetic spin-polarized calculations were initially performed 
for each of the 20 different dopant configurations.  Low-energy configurations were identified from these results (Fig.\ref{fig1}(2), Fig.\ref{fig2}(4), 
Fig.\ref{fig3}(2), and Fig.\ref{fig4}(1) for systems with 2, 3, 4 and 5 Cr atoms respectively.  Fig.\ref{fig2}(3) instead of Fig.\ref{fig2}(4) has low energy 
with Ru and Rh dopants.  Other configurations are same).  With these lowest-energy configurations, subsequent calculations were performed for the 
ferromagnetic (FM) and antiferromagnetic (AFM) spin states of the dopants in their respective charge states (controlled by specifying the NELECT and 
NUPDOWN tags in the INCAR file).

This work predicts 4.2 eV for the band gap of w-AlN.  This is in good agreement with the 4.054 eV from the Materials Project database\cite{mp661}.  
Experimental band gap is 6.2 eV\cite{strite1992}.  GGA generally underestimate band gaps\cite{grumet2018}.  Nevertheless, we used the PBE/GGA functional in 
this work because recent studies on transition metal-doped AlN and GaN report that, compared to PBE, both the hybrid functional and the Hubbard U approach 
worsen the agreement of defect levels with experimental data.  This occurs even though the hybrid functional can reproduce the experimental band gap 
through empirical tuning of the exchange fraction\cite{zakrzewski2016,schultz2023}.  John-Teller distortion is reported to be too small to significantly 
affect magnetic interactions in Mn-doped GaN at the doping concentration relevant to magnetic semiconductors\cite{sato2010}.  We assume similar trend 
for Cr-, Ru- and Rh-doped AlN and, therefore, omit analysis of Jahn-Teller distortions in our optimized structures.

\section{\label{results}RESULTS AND DISCUSSION}
\subsection{\label{thermo}Formation energies and oxidation states of dopants}
\begin{figure}
\includegraphics[width=0.48\textwidth]{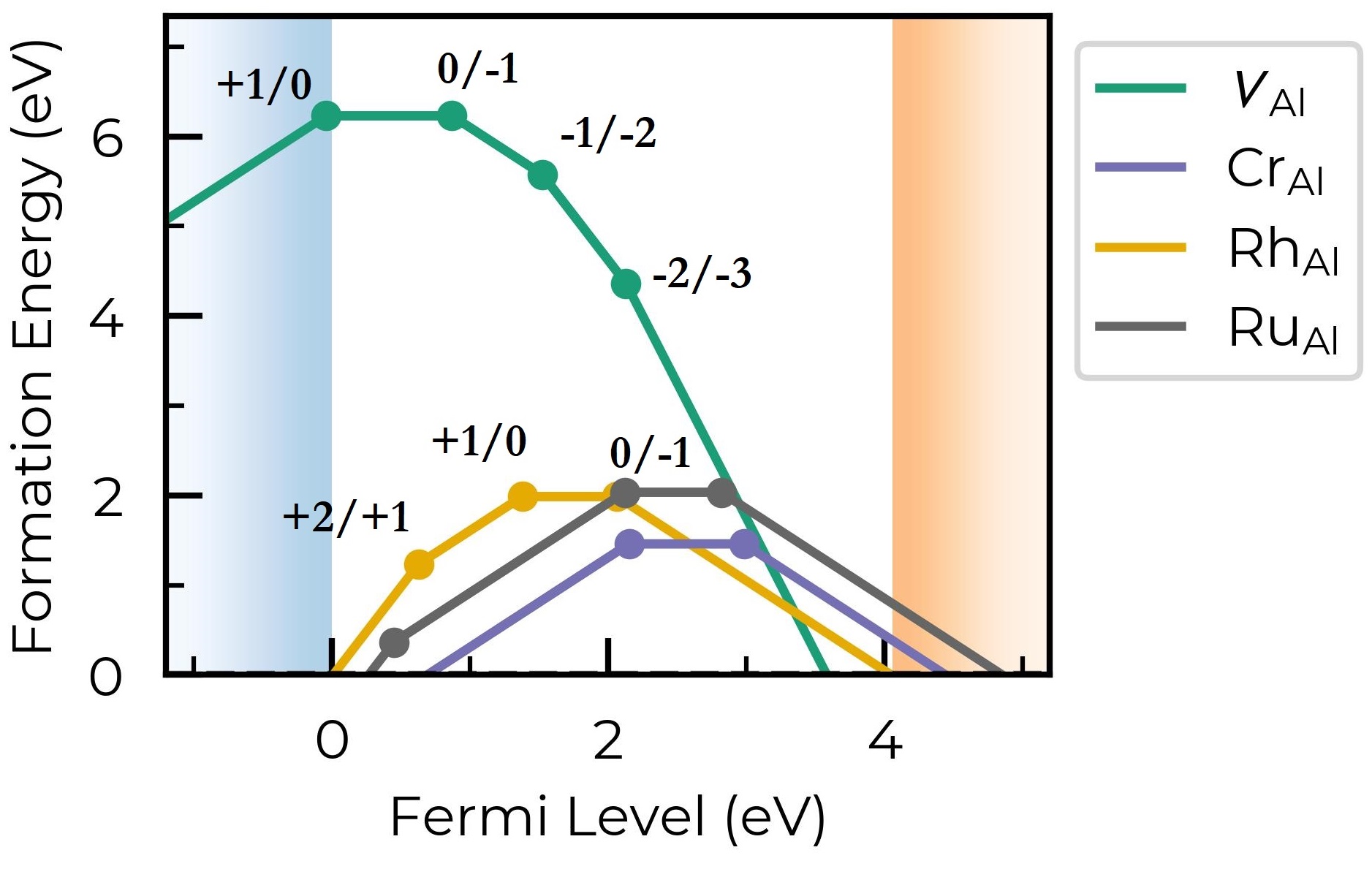}
\caption{\label{fig5}Defect formation energies for point defects, namely, aluminum vacancy ($V_{\mathrm{Al}}$), substitutional chromium ($\mathrm{Cr_{Al}}$),
substitutional ruthenium ($\mathrm{Ru_{Al}}$), and substitutional rhodium ($\mathrm{Rh_{Al}}$) in w-AlN as a function of the Fermi level under nitrogen-rich 
condition.}
\end{figure}

The computed equilibrium lattice parameters of w-AlN ($a$ = 3.11 $\text{\AA{}}$ and $c$ = 5.0 $\text{\AA{}}$) are in excellent agreement with the 
experimental values of 3.11 $\text{\AA{}}$ and 4.98 $\text{\AA{}}$\cite{strite1992}.  Figure~\ref{fig5} presents the calculated point-defect formation 
energies as a function of the Fermi level $E_F$ in w-AlN for Nitrogen-rich condition.  The point defects considered in this work include the aluminum 
vacancy ($V_{\mathrm{Al}}$), substitutional chromium ($\mathrm{Cr_{Al}}$), substitutional ruthenium ($\mathrm{Ru_{Al}}$), and substitutional rhodium 
($\mathrm{Rh_{Al}}$).  To keep the $E^f(X^q)$ versus $E_F$ plot clear, the full straight line for every charge state of each defect is not shown.  
Instead, only the section where a given charge state has a lower energy than all other charge states is displayed.  Consequently, a change in the slope 
of these lines indicates a transition in the charge state of the defect\cite{prashun2025}.  The formation energy of $V_{\mathrm{Al}}$ as a function of Fermi 
energy is in excellent agreement with the literature\cite{stampfl2002}.  The minor differences are likely due to differences in the structure and the 
exchange-correlation interaction approximation.  With these formation energies, the maximum equilibrium concentration of $V_{\mathrm{Al}}$ at the 
melting point of AlN (2470 K) is approximately $2.4\times10^{-10}$ percent.  This concentration is insufficient to produce any detectable magnetism 
by vacancy induced spin polarization\cite{osorio2006}.

This work predicts that Cr$^{4+}$ and Ru$^{4+}$ are the most probable charge states for the $\mathrm{Cr_{Al}}$ and $\mathrm{Ru_{Al}}$ defects.  For the 
$\mathrm{Rh_{Al}}$ defect, Rh$^{3+}$ and Rh$^{2+}$ are predicted to be the likely charge states, with populations of 65\% and 35\% respectively. 
Our predicted stable charge state of Cr$^{4+}$ for $\mathrm{Cr_{Al}}$ is in agreement with the photoluminescence studies\cite{baur1995}.  The formation 
energies of the Cr$^{4+}$, Ru$^{4+}$, Rh$^{3+}$, and Rh$^{2+}$ defects are computed to be 1.34 eV, 1.94 eV, 1.98 eV and 2.02 eV respectively.  To the best 
of our knowledge, there is no existing literature on the formation energies of these specific point defects in w-AlN as a function of Fermi level.  At the 
melting point of AlN, these defect formation energies yield equilibrium concentrations of 0.092\%, 0.006\%, and 0.04\% for the Cr$^{4+}$, Ru$^{4+}$, and 
Rh$^{3+}$ ions, respectively.  This indicates that doping w-AlN sufficiently to support dilute magnetism under equilibrium conditions is difficult; 
hence, non-equilibrium growth techniques, such as molecular beam epitaxy, would be required\cite{frazier2005}.

\begingroup
\begin{table*}[htbp]
\caption{Magnetic polarization of dopant ions (Cr$^{4+}$, Ru$^{4+}$, and Rh$^{3+}$) in different configurations within 3$\times$3$\times$3 w-AlN 
supercells.  Column 1 lists the doped supercell systems.  The energy differences between the ferromagnetic (FM) and the antiferromagnetic (AFM) states 
($\Delta$E$_{\text{FM-AFM}}$) are listed in column 2.  Columns 3 and 4 list the relaxed FM and AFM spin magnetic moments on the dopant atoms.}
\label{table1}
\resizebox{\textwidth}{!}{
\begin{tabular}{l|l|l|l}
\hline \hline
System                  &$\Delta$E$_{\text{FM-AFM}}$ (eV/atom)&FM Spin moments ($\mu_B$)&AFM Spin moments ($\mu_B$)\\\hline
Al$_{53}$Cr$_1$N$_{54}$ &-0.009&+1.886                                                         &0.000\\
Al$_{52}$Cr$_2$N$_{54}$ &-0.001&+1.926 +1.926                                                  &-1.830 +1.830\\
Al$_{51}$Cr$_3$N$_{54}$ &-0.008&+1.934 +1.934 +1.934                                           &+1.263 -2.435 +1.255\\
Al$_{50}$Cr$_4$N$_{54}$ &-0.009&+2.006 +2.010 +1.900 +1.884                                    &+0.084 +0.084 +2.018 -2.090\\
Al$_{49}$Cr$_5$N$_{54}$ &-0.020&+1.880 +2.014 +2.022 +1.860 +2.112                             &-0.170 +0.342 +0.356 -0.186 -0.270\\
Al$_{47}$Cr$_7$N$_{54}$ &-0.023&+1.999 +1.998 +2.014 +1.976 +2.220 +1.910 +1.904               &-0.426 -0.359 -0.365 -0.435 +2.377 -0.413 -0.427\\
Al$_{45}$Cr$_9$N$_{54}$ &-0.042&+1.997 +1.997 +1.997 +1.997 +1.997 +1.997 +1.997 +1.997 +1.997 &0.000 0.000 0.000 0.000 0.000 0.000 0.000 0.000\\
Al$_{53}$Ru$_1$N$_{54}$ &0.012&+2.452                                                          &0.000\\
Al$_{52}$Ru$_2$N$_{54}$ &0.026&+2.441 +2.440                                                   &0.000 0.000\\
Al$_{51}$Ru$_3$N$_{54}$ &0.041&+2.508 +2.356 +2.356                                            &0.000 0.000 0.000\\
Al$_{50}$Ru$_4$N$_{54}$ &0.061&+2.195 +2.193 +2.683 +2.351                                     &0.000 0.000 0.000 0.000\\
Al$_{49}$Ru$_5$N$_{54}$ &0.092&+2.250 +2.738 +2.734 +2.247 +2.050                              &0.000 0.000 0.000 0.000 0.000\\
Al$_{47}$Ru$_7$N$_{54}$ &0.142&+2.482 +2.488 +2.476 +2.488 +2.038 +2.542 +2.538                &0.000 0.000 0.000 0.000 0.000 0.000 0.000\\
Al$_{45}$Ru$_9$N$_{54}$ &0.208&+2.540 +2.540 +2.540 +2.540 +2.540 +2.540 +2.540 +2.540 +2.540  &0.000 0.000 0.000 0.000 0.000 0.000 0.000 0.000\\
Al$_{53}$Rh$_1$N$_{54}$ &-0.002&+1.070                                                         &0.000\\
Al$_{52}$Rh$_2$N$_{54}$ & 0.006&+1.044 +1.040                                                  &0.000 0.000\\
Al$_{51}$Rh$_3$N$_{54}$ & 0.005&+1.028 +0.982 +1.016                                           &0.000 0.000 0.000\\
Al$_{50}$Rh$_4$N$_{54}$ & 0.016&+0.885 +0.878 +1.363 +1.193                                    &0.000 0.000 0.000 0.000\\
Al$_{49}$Rh$_5$N$_{54}$ & 0.015&+0.952 +1.328 +1.316 +0.960 +0.850                             &0.000 0.000 0.000 0.000 0.000\\
Al$_{47}$Rh$_7$N$_{54}$ & 0.039&+1.140 +1.136 +1.112 +1.169 +0.800 +1.232 +1.236               &0.000 0.000 0.000 0.000 0.000 0.000 0.000\\
Al$_{45}$Rh$_9$N$_{54}$ & 0.064&+1.173 +1.173 +1.173 +1.173 +1.173 +1.173 +1.173 +1.173 +1.173 &0.000 0.000 0.000 0.000 0.000 0.000 0.000 0.000\\\hline \hline
\end{tabular}
}
\end{table*}
\endgroup

\subsection{\label{mag}Magnetic properties}
As discussed previously, the energy difference between the ferromagnetic (FM) and antiferromagnetic (AFM) states is used to determine the stable 
magnetic ground state of transition metal doped semiconductors.  Therefore, we carried out FM and AFM spin state calculations for the selected 
low-energy configurations of the AlN supercell at each of the seven concentrations of each dopant (Cr$^{4+}$, Ru$^{4+}$, and Rh$^{3+}$).  
Note that the FM simulation sets the total spin magnetization to the number of unpaired d electrons (in $\mu_B$) from the transition metal dopants, 
whereas the AFM simulation constrains it to zero.  Table~\ref{table1} collects the results of these calculations.  A negative value of the energy difference, 
$\Delta$E$_{\text{FM-AFM}}$, means that the FM state is more stable than the AFM state.  It is evident from the energy differences in column 2 that the FM state is 
favored for all considered concentrations of Cr$^{4+}$.  However, for Ru$^{4+}$- and Rh$^{3+}$- doped systems, the AFM states are favored over the FM states, except 
for the system with a single Rh$^{3+}$ ion.  Furthermore, it is evident that $\Delta$E$_{\text{FM-AFM}}$ is approximately an order of magnitude greater for Ru-doped 
systems than for Cr- and Rh-doped systems.  

As stated earlier, Cr$^{4+}$, Ru$^{4+}$, and Rh$^{3+}$ (with 3d$^2$, 4d$^4$, and 4d$^6$ residual valence electron configurations, respectively) are the 
most likely charge states for $\mathrm{Cr_{Al}}$, $\mathrm{Ru_{Al}}$ and $\mathrm{Rh_{Al}}$ defects.  The tetrahedral crystal field of the surrounding 
ligands splits the five fold degenerate d states of the free transition metal atoms into two low lying e (d$_{z^2}$ and d$_{x^2-y^2}$) and three high 
lying t$_2$ (d$_{xy}$, d$_{xz}$, and d$_{yz}$) subsets.  The transition metal ions with the configuration d$^n$ (n=3 to 6) could, in principle, adopt 
either a low-spin (LS) or a high-spin (HS) state, depending on the competition between the crystal-field splitting energy and the spin-pairing energy.  
A large crystal-field splitting energy favors the LS state, whereas a large spin-pairing energy favors HS state\cite{sato2010}.

The stability of the FM state over the AFM state for Cr-doped w-AlN systems can be understood as follows.  It is evident from Table~\ref{table1} that 3d$^2$ 
electrons of Cr$^{4+}$ ions occupy the lower-lying e orbitals with unpaired parallel spins, e$^2$($\uparrow\uparrow$) t$_2^0$, yielding a local spin moment of 
about 2$\mu_B$ on each Cr$^{4+}$ ion in all configurations.  Because Cr$^{4+}$ has a d$^{2}$ configuration, its spin alignment is governed strictly by Hund's rule 
within the e orbitals, rather than a competition between crystal-field splitting and spin-pairing energies.  This robust local parallel spin alignment facilitates a 
strong ferromagnetic exchange interaction between the dopant ions, ultimately favoring the stability of the FM state over the AFM state.  

The stability of the AFM state over the FM state for Ru-doped systems shows that the d-electrons of Ru$^{4+}$ ions adopts a low-spin state, 
e$^4$($\uparrow\downarrow,\uparrow\downarrow$) t$_2^0$.  This electronic configuration leads to a fully paired, net zero local spin magnetic moment 
on each individual Ru atom.  Similarly, for the Rh-doped systems, the stability of the AFM state indicates that the Rh$^{3+}$ ions favor a completely 
spin-paired low-spin configuration.  While an ideal tetrahedral low-spin d$^6$ configuration would usually leave two unpaired electrons in the t$_2$ level 
(e$^4(\uparrow\downarrow,\uparrow\downarrow$) t$_2^2(\uparrow\uparrow$)), the calculated local magnetic moments of 0$\mu_B$ on all Rh$^{3+}$ ions across 
all concentrations imply a strong local structural distortion or significant hybridization.  This splits the t$_2$ states and forces the remaining electrons 
to fully pair up into a non-magnetic state (e$^4(\uparrow\downarrow,\uparrow\downarrow$) t$_2^2(\uparrow\downarrow$)).  For both dopants, imposing a high-spin 
state costs energy, confirming that they behave as locally non-magnetic ions stabilized in their respective low-spin configurations.

\begin{figure}
\includegraphics[width=0.45\textwidth]{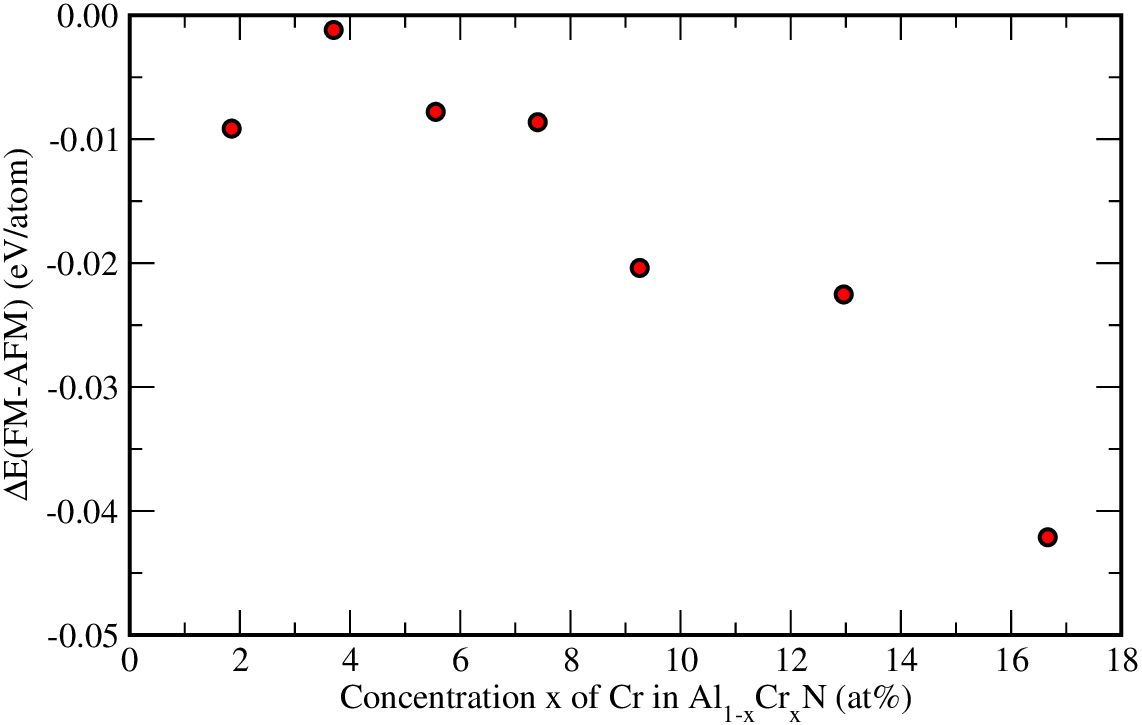}
\caption{\label{fig6}The total energy difference between the ferromagnetic and antiferromagnetic states of Cr-doped w-AlN, $\Delta$E$_{\text{FM-AFM}}$, 
as a function of Cr concentration.}
\end{figure}

\begin{figure}
\includegraphics[width=0.45\textwidth]{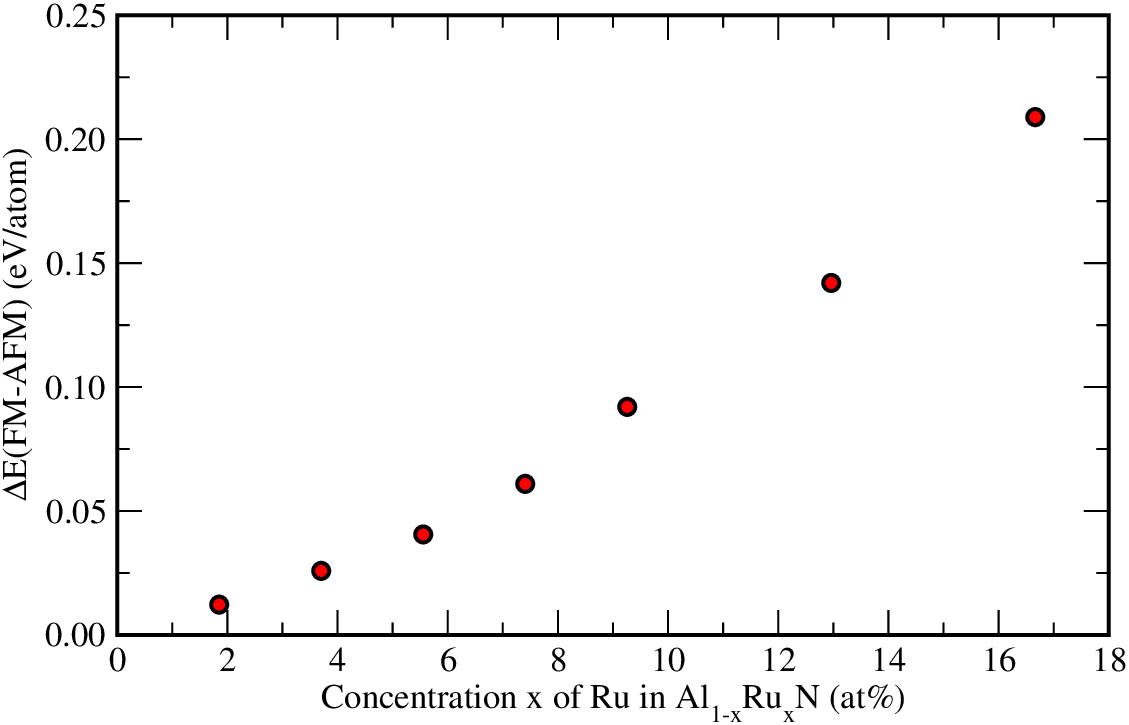}
\caption{\label{fig7}The total energy difference between the ferromagnetic and antiferromagnetic states of Ru-doped w-AlN, $\Delta$E$_{\text{FM-AFM}}$, 
as a function of Ru concentration.}
\end{figure}

\begin{figure}
\includegraphics[width=0.45\textwidth]{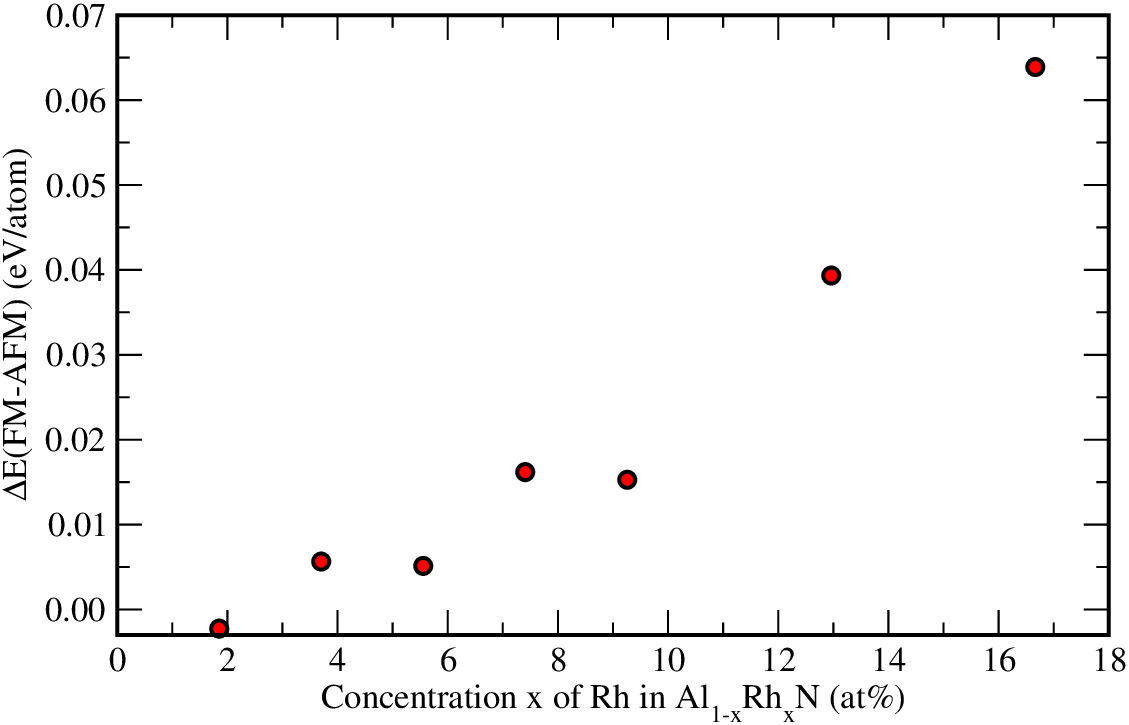}
\caption{\label{fig8}The total energy difference between the ferromagnetic and antiferromagnetic states of Rh-doped w-AlN, $\Delta$E$_{\text{FM-AFM}}$, 
as a function of Rh concentration.}
\end{figure}

Figure~\ref{fig6} demonstrates that the FM state is consistently more stable than the AFM state for Cr$^{4+}$-doped AlN supercells across seven concentrations 
up to 16.67\% Al substitution, a behavior driven by the d$^{2}$ electronic configuration and Hund's rule.  While this FM stability strengthens with increased 
Cr concentration, the Ru- and Rh-doped systems exhibit a sharp contrast as shown in Figs.~\ref{fig7} and~\ref{fig8}.  For Ru, the low-spin configuration of 
Ru$^{4+}$ favors the AFM state over the FM state across all considered concentrations.  Similarly, the spin-paired non-magnetic state of Rh stabilizes the AFM 
phase at all but the lowest concentration (a single Rh substitution in Al$_{54}$N$_{54}$).  Consequently, the energy difference, $\Delta$E$_{\text{FM-AFM}}$, 
becomes increasingly positive with higher concentrations of Ru and Rh dopants, behaving inversely to the concentration-dependent stability observed in the
Cr-doped system.

Regarding computational accuracy, the PBE functional remains a standard general purpose GGA for solid-state properties, offering a balanced performance in 
predicting structure and energetics\cite{burke2016,zhang2018}.  Its mean absolute error for compound formation enthalpy is 0.052 eV/atom, which establishes 
the threshold for chemical accuracy.  For the 16.67\% Cr-doped AlN system, the calculated $\Delta$E$_{\text{FM-AFM}}$ is -0.042 eV/atom.  Because this 
energy difference is on the order of the functional's chemical accuracy, the FM state in w-AlN is expected to be stable and experimentally sustainable.

\subsection{\label{dos}Electronic density of states}
\begin{figure*}
\includegraphics[width=0.85\textwidth]{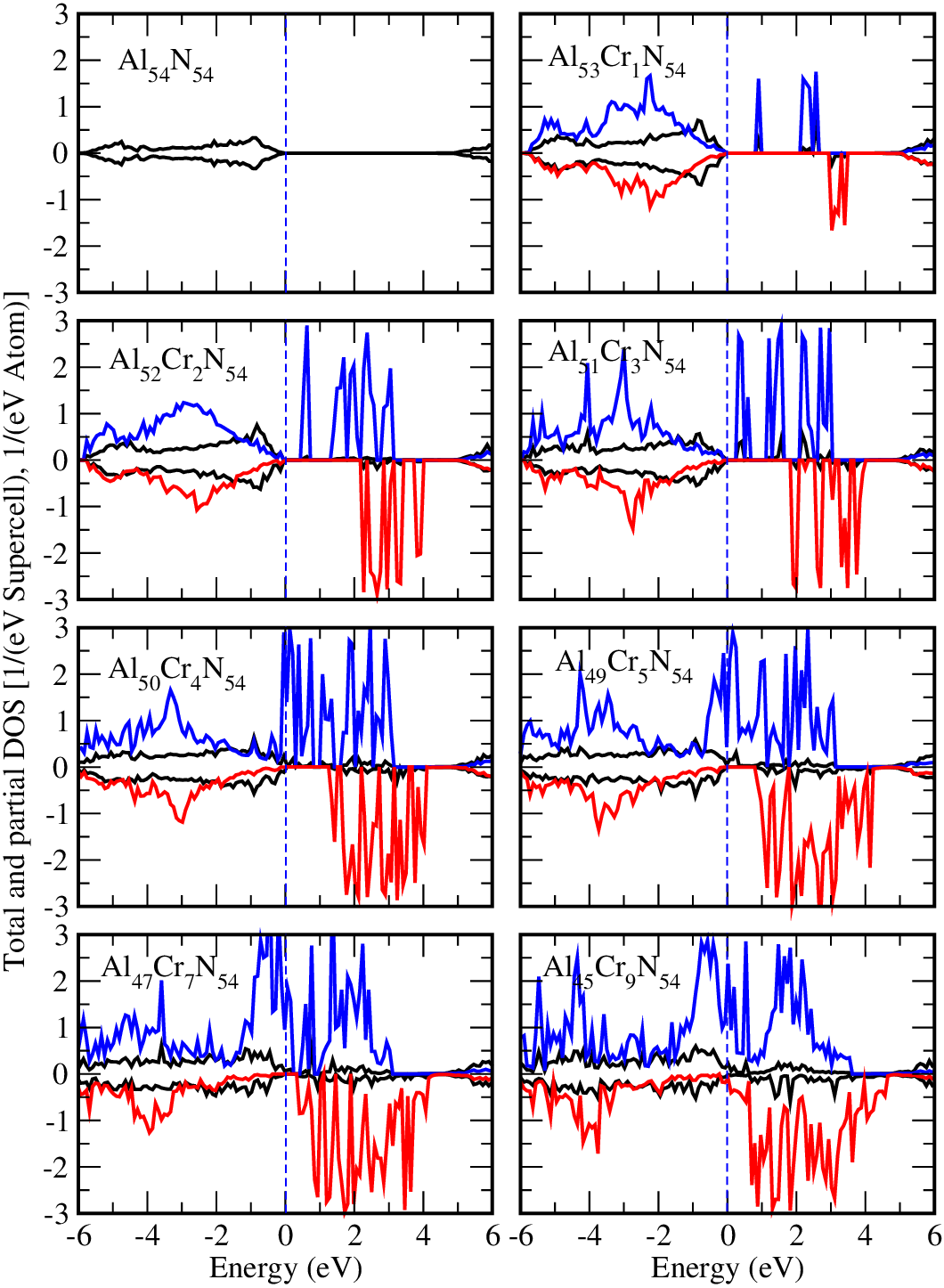}
\caption{\label{fig9}Electronic density of states of w-AlN with Cr dopant atoms in the ferromagnetic spin state, modelled using a 3$\times$3$\times$3 supercell.  
Black lines show the total spin-up and spin-down DOS per supercell.  Blue and red lines show the respective spin-up and spin-down 3d partial DOS per Cr atom, 
averaged over all dopant atoms in the supercell.  The graph labeled Al$_{54}$N$_{54}$ shows the total density of states of pure w-AlN.  The vertical line at 
0 eV represents the Fermi level.}
\end{figure*}

\begin{figure*}
\includegraphics[width=0.85\textwidth]{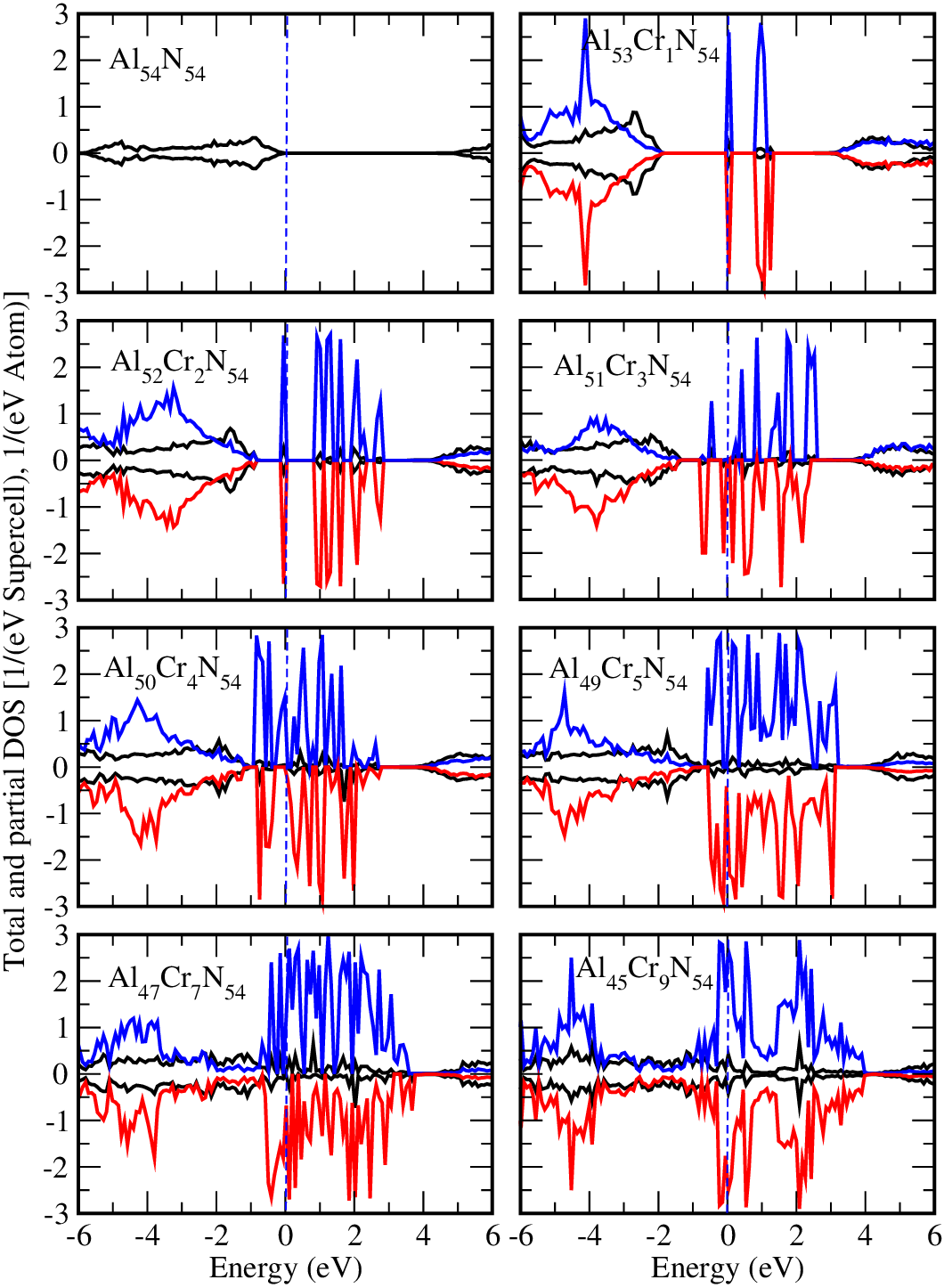}
\caption{\label{fig10}Electronic density of states of w-AlN with Cr dopant atoms in the antiferromagnetic spin state, modelled using a 3$\times$3$\times$3 supercell.  
Black lines show the total spin-up and spin-down DOS per supercell.  Blue and red lines show the respective spin-up and spin-down 3d partial DOS per Cr atom, 
averaged over all dopant atoms in the supercell.  The graph labeled Al$_{54}$N$_{54}$ shows the total density of states of pure w-AlN.  The vertical line at 
0 eV represents the Fermi level.}
\end{figure*}

\begin{figure*}
\includegraphics[width=0.85\textwidth]{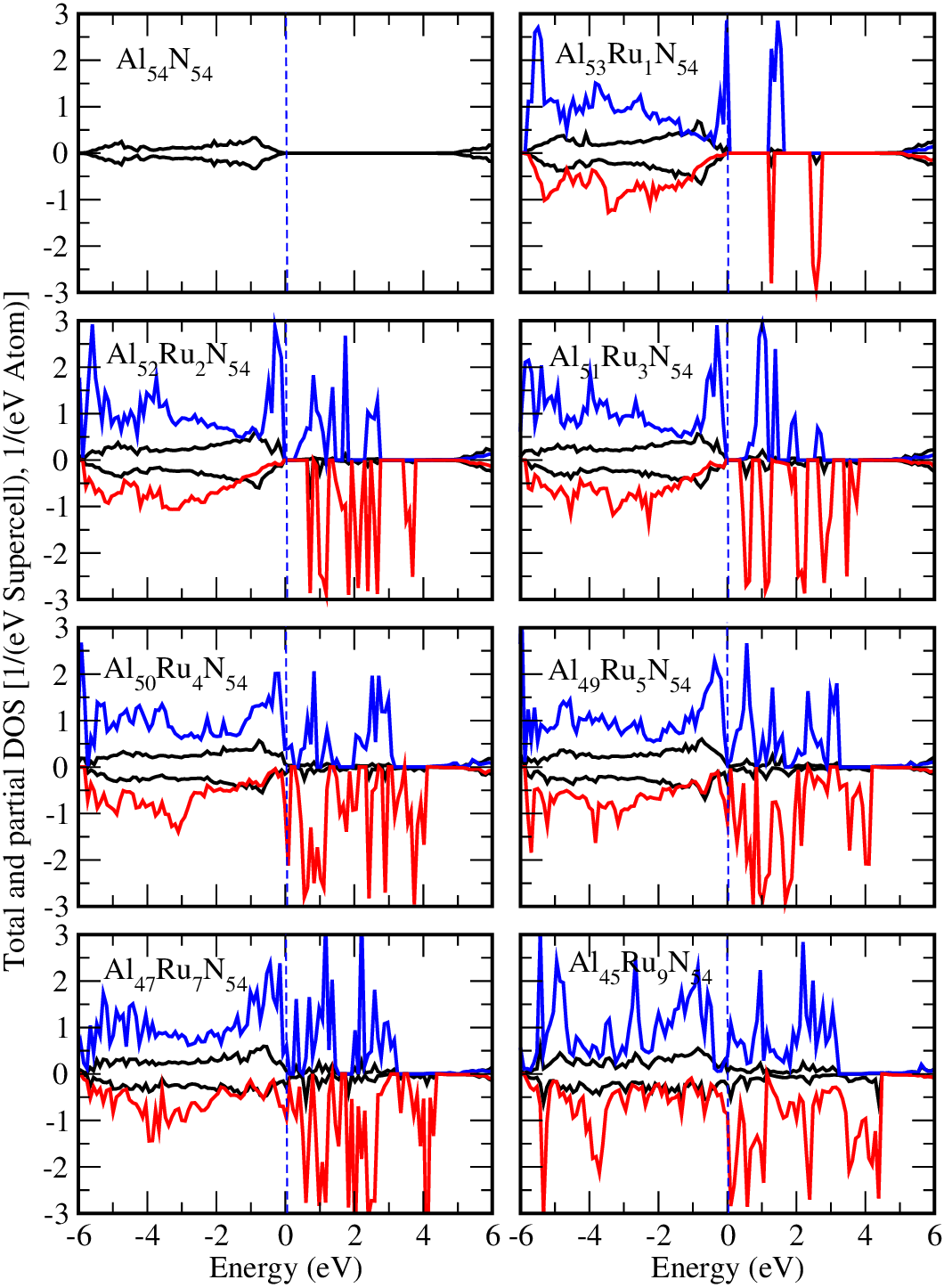}
\caption{\label{fig11}Electronic density of states of w-AlN with Ru dopant atoms in the ferromagnetic spin state, modelled using a 3$\times$3$\times$3 supercell.  
Black lines show the total spin-up and spin-down DOS per supercell.  Blue and red lines show the respective spin-up and spin-down 4d partial DOS per Ru atom, 
averaged over all dopant atoms in the supercell.  The graph labeled Al$_{54}$N$_{54}$ shows the total density of states of pure w-AlN.  The vertical line at 
0 eV represents the Fermi level.}
\end{figure*}

\begin{figure*}
\includegraphics[width=0.85\textwidth]{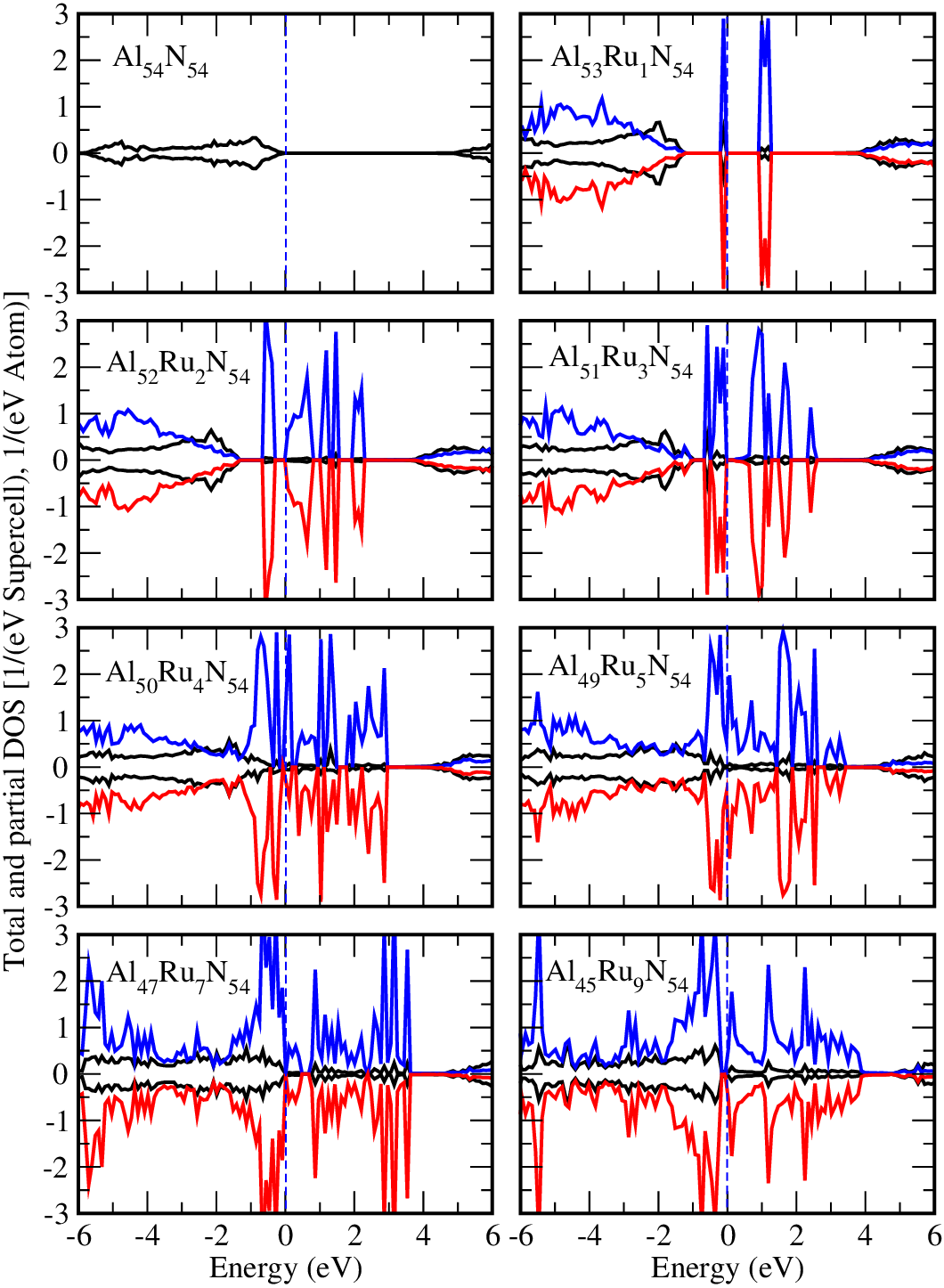}
\caption{\label{fig12}Electronic density of states of w-AlN with Ru dopant atoms in the antiferromagnetic spin state, modelled using a 3$\times$3$\times$3 supercell.  
Black lines show the total spin-up and spin-down DOS per supercell.  Blue and red lines show the respective spin-up and spin-down 4d partial DOS per Ru atom, 
averaged over all dopant atoms in the supercell.  The graph labeled Al$_{54}$N$_{54}$ shows the total density of states of pure w-AlN.  The vertical line at 
0 eV represents the Fermi level.}
\end{figure*}

\begin{figure*}
\includegraphics[width=0.85\textwidth]{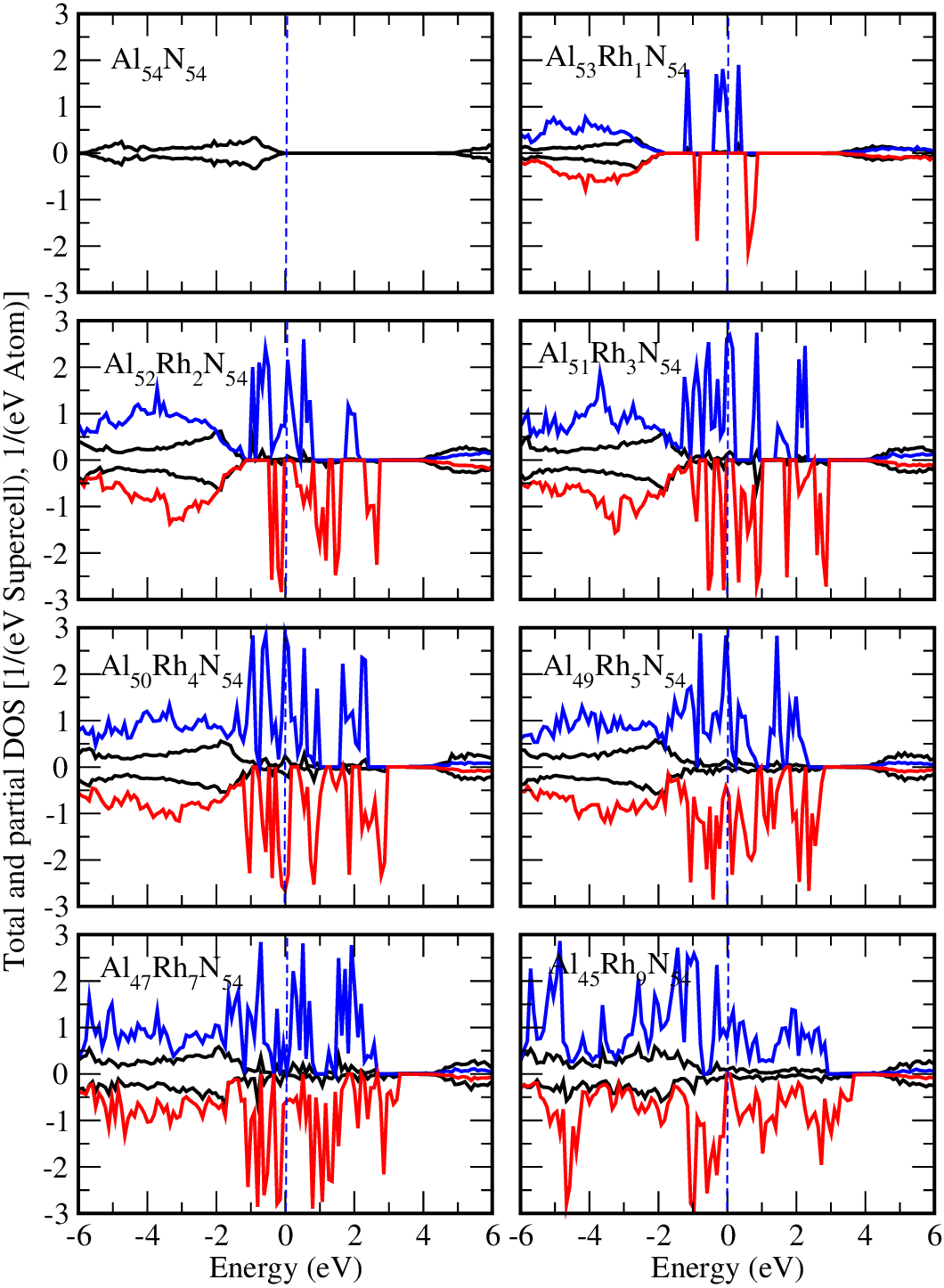}
\caption{\label{fig13}Electronic density of states of w-AlN with Rh dopant atoms in the ferromagnetic spin state, modelled using a 3$\times$3$\times$3 supercell.  
Black lines show the total spin-up and spin-down DOS per supercell.  Blue and red lines show the respective spin-up and spin-down 4d partial DOS per Rh atom, 
averaged over all dopant atoms in the supercell.  The graph labeled Al$_{54}$N$_{54}$ shows the total density of states of pure w-AlN.  The vertical line at 
0 eV represents the Fermi level.}
\end{figure*}

\begin{figure*}
\includegraphics[width=0.85\textwidth]{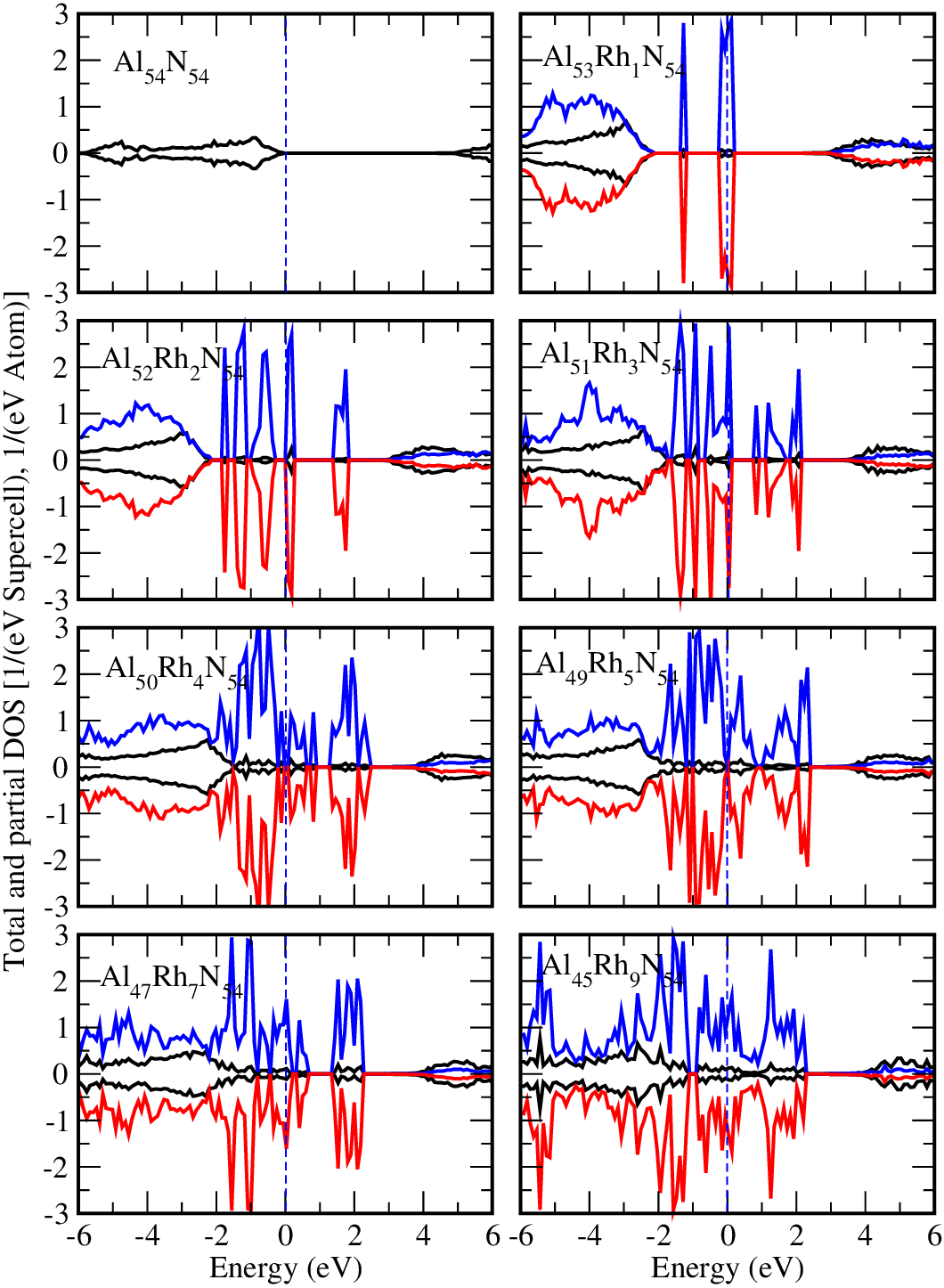}
\caption{\label{fig14}Electronic density of states of w-AlN with Rh dopant atoms in the antiferromagnetic spin state, modelled using a 3$\times$3$\times$3 supercell.  
Black lines show the total spin-up and spin-down DOS per supercell.  Blue and red lines show the respective spin-up and spin-down 4d partial DOS per Rh atom, averaged 
over all dopant atoms in the supercell.  The graph labeled Al$_{54}$N$_{54}$ shows the total density of states of pure w-AlN.  The vertical line at 0 eV represents 
the Fermi level.}
\end{figure*}

This work examines the possibility of ferromagnetic state in w-AlN doped with Cr, Ru and Rh atoms.  It is seen from the energy differences  that Cr doped AlN 
supports the FM state over the AFM state, whereas Ru and Rh doped AlN favor the AFM state.  The electronic density of states (DOS) of these systems are compared here.  
Figures \ref{fig9} and \ref{fig10} present the ferromagnetic (FM) and antiferromagnetic (AFM) density of states (DOS) for Cr$^{4+}$-doped w-AlN over seven 
different concentrations: Al$_{54-x}$Cr$_x$N$_{54}$, where x = 1, 2, 3, 4, 5, 7, and 9.  Corresponding DOS graphs of Ru$^{4+}$-doped w-AlN are shown in Figs.\ref{fig11} 
and \ref{fig12}, while those of Rh$^{3+}$-doped systems are shown in Figs.\ref{fig13} and \ref{fig14}.  Note that the FM simulation sets the total spin magnetization 
to the number of unpaired d electrons (in $\mu_B$) from the transition metal dopants, whereas the AFM simulation constrains it to zero.

It is evident from the DOS graphs of Cr doped AlN (Figs. \ref{fig9} and \ref{fig10}) that the d orbitals from the substitutional dopant Cr hybridizes well 
with the s and p orbitals of neighboring nitrogen (N) ligands.  Comparison of the AFM and FM DOS reveals that the valence states in the FM configuration are 
significantly broadened and pushed deeper into the valence band, down to -6 eV, whereas they are restricted to the -5 to 0 eV range in the AFM configuration.  
This deeper, broader positioning of the valence states maximizes bonding interactions, thereby stabilizing the FM state.  This observation corroborates the energy 
differences presented in Table~\ref{table1} and Figure~\ref{fig6}.

The FM DOS graphs in Fig.\ref{fig9} show that the spin-up and spin-down states are highly asymmetric for all Cr concentrations.  This asymmetry is driven by 
two competing quantum mechanical effects:  Tetrahedral crystal field splits the Cr 3d orbitals into a lower-energy e doublet and a higher-energy t$_2$ triplet.  
A Cr$^{4+}$ ion has a 3d$^2$ configuration.  Guided by Hund's rule, both electrons occupy the spin-up e sub-band, making it full, while the spin-up t$_2$ 
and all spin-down sub-bands remain empty (e$^2_{\uparrow}$ t$^0_{2\uparrow}$ e$^0_{\downarrow}$ t$^0_{2\downarrow}$).  Exchange splitting pushes the entire spin-down 
channel (e$^0_{\downarrow}$, t$^0_{2\downarrow}$) to higher, unoccupied energies.  The density of states due to the two electrons from the single Cr$^{4+}$ ion in 
Al$_{53}$Cr$_1$N$_{54}$ (i.e., at x = 1) appears as a small blue peak or kink on the host valence band edge, just below the Fermi level.  Because the e$_{\uparrow}$ 
band is full and the higher t$_{2\uparrow}$ band is empty, the Fermi level (0 eV) passes through a local gap, making the system a ferromagnetic semiconductor.  

For x = 2, all four electrons from the  Cr$^{4+}$ ions occupy the combined e$_{\uparrow}$ states.  Interactions between the Cr atoms 
cause these states to broaden, making the blue kink grow into an enhanced peak.  Because the e$_{\uparrow}$ manifold has a total capacity of four electrons for 
two atoms, it remains exactly filled.  At this low concentration, the Fermi level still passes through a local gap directly above this filled e$_{\uparrow}$ band.

As the doping level increases, the DOS graphs show that the system transitions through three electronic phases:  For 
x = 1 to 3, the e$_{\uparrow}$ impurity states are localized and remain completely filled; because they are located within the host valence band, the system remains 
in the ferromagnetic semiconductor phase.  Enhanced Cr-N-Cr hybridization causes considerable broadening of the 3d bands for x = 4 to 7.  The filled e$_{\uparrow}$ 
band and the empty t$_{2\uparrow}$ band begin to overlap, causing the spin-up channel to cross the Fermi level.  The spin-down channel remains completely empty 
at the Fermi level, achieving 100\% spin polarization and turning the system into a half-metallic ferromagnetic phase.  At very high doping concentration (x = 9), 
crowding and strong interactions among the dopants shift the bands substantially. The exchange splitting is no longer enough to keep the spin-down states entirely 
empty.  The Fermi level eventually cuts through both the spin-up and spin-down Cr 3d states, converting the half-metal state into a normal metal state.  

The AFM DOS graphs in Fig.\ref{fig10} show that the spin-up and spin-down DOS are symmetric for x = 2 and 9, but asymmetric for x = 1, 3, 4, 5, and 7.  
Enforcing a singlet (antiparallel) spin configuration on a single Cr$^{4+}$ ion imposes a low-spin state on an ion that naturally prefers a high-spin 
triplet (parallel) state.  The apparent symmetry in the occupied part of the associated DOS confirms that imposing the low-spin state was successful; however, 
because this violates Hund’s rule of maximum multiplicity, the high-spin state remains energetically favorable due to exchange energy gain-as reflected in the 
$\Delta$E$_{\text{FM-AFM}}$ values in Table~\ref{table1}.  Notice the sharp peaks located precisely at the Fermi level for x = 1 while in the corresponding FM DOS, 
the Fermi level is located within a local gap.  According to the Stoner criterion, a high density of states at the Fermi level creates an electronic instability.  
The system resolves this electronic instability by undergoing exchange splitting (spontaneous spin polarization), which shifts the states and lowers the total 
energy, making FM state favored.  For x = 2, the four 3d electrons are distributed such that each Cr ion maintains a high-spin state with opposite polarization 
(e$^2(\uparrow\uparrow$), e$^2(\downarrow\downarrow$)) as is evident from the local spin moments of $\pm$1.83$\mu_B$ (Table~\ref{table1}).  The DOS is symmetric 
because the majority states of one Cr ion exactly mirror the minority states of the other Cr ion.  Notice the sharp peaks located precisely at the Fermi level 
while in the corresponding FM DOS, the Fermi level is located within a local gap.  This indicates that the inter-atomic ferromagnetic coupling provides a 
greater total energy reduction than the AFM interaction, though the intra-atomic exchange energy is restored which was lost in the x=1 case.  

The convergence of the local magnetic moments on all nine Cr$^{4+}$ ions to exactly 0.000—despite a ferromagnetic initialization—indicates a complete spin 
quenching driven by the constraint NUPDOWN = 0.  Because the nine (x = 9) Cr$^{4+}$ dopants occupy nearest-neighbor Al sites on the a-b plane of the w-AlN lattice, 
their 3d orbitals overlap both directly Cr\textendash Cr and indirectly through the bridging nitrogen ligands Cr\textendash N\textendash Cr. This spatial overlapping 
causes the individual atomic d-electrons to delocalise entirely across the 2D cluster, turning into a broad band rather than remaining localized on isolated atomic 
centers.  Consequently, under the net-zero spin constraint, the system minimizes energy by symmetrically populating each delocalized band state with an identical 
number of spin-up and spin-down electrons. This results in a delocalized, non-magnetic metallic state rather than a collection of individual atomic singlets.

For x = 3, 4, 5, and 7, the Cr atoms tend to be magnetic, like x = 1 case, but they cannot achieve a standard ferromagnetic state because the NUPDOWN = 0 forbids 
the cell from having a net magnetic moment.  To satisfy both conditions (remaining locally magnetic while forcing a total net moment of zero), the system finds a 
complex antiferromagnetic solution.  Take x = 3 for example, the local moments are: +1.26, -2.43, +1.26.  The sum of these moments (1.26 + 1.26 - 2.43 = +0.09) is 
effectively zero.  Consequently, even though the total area under the spin-up and spin-down DOS curves is equal, the individual shapes and energy positions of the 
spin channels are asymmetric.  This occurs because the individual Cr ions are in different local environments or posses unequal spin moments (e.g., one Cr atom has a 
moment of -2.43, while the other two have +1.26).  The exchange splitting experienced by an electron depends on its local atomic environment, which shifts the peaks 
unevenly and result in an asymmetric visual profile despite having equal total integrated areas.

The electronic density of states of Ru$^{4+}$-doped w-AlN, Figs.\ref{fig11} and \ref{fig12}, is examined next.  It is evident from Table~\ref{table1} 
that the AFM state is favored over the FM state for all dopant concentrations (x = 1, 2, 3, 4, 5, 7, and 9).  The local tetrahedral symmetry of the w-AlN host 
splits the 4d orbitals of the Ru$^{4+}$ (4d$^{4}$) ions into lower-energy e and higher-energy t$_{2}$ subshells, analogous to the Cr-doped system.
The resulting electron occupation within these split levels determines the overall stability of the system under different spin constraints.

Under the FM spin specification (NUPDOWN = ND, where ND is total number of d-electrons), the electronic structure breaks spin symmetry uniformly across 
the lattice.  Driven by Hund's rule and dominant exchange interactions, every individual Ru$^{4+}$ ion adopts an identical high-spin alignment, resulting 
in an e$^2$($\uparrow\uparrow$) t$_2^2$($\uparrow\uparrow \cdot$) configuration on each dopant site.  This uniform polarization yields a consistent 
local spin magnetic moment of approximately +2.45$\mu_B$ per Ru atom.

However, as shown in the FM density of states (Fig.~\ref{fig11}), this high-spin configuration alters the electronic landscape near the Fermi level 
(E$_F$ = 0 eV) depending on the concentration.  For the single-dopant case (x = 1), a sharp, Ru 4d impurity peak is placed directly at E$_{F}$ within the 
majority spin channel, inducing a kinetic energy penalty.  At intermediate concentrations (x = 2 and 3), E$_{F}$ falls within a local energy gap; yet, this 
configuration remains energetically unfavorable, indicating that the exchange energy gained by separating the spin channels is inadequate to offset the 
structural and electronic relaxation energy available in the spin-paired state.  At higher concentrations (x=4, 5, 7, and 9), the system transitions into a 
conventional metallic state featuring a finite, continuous density of states rather than isolated peaks.  According to the Stoner criterion ($I \cdot N(E_F) > 1$), 
spontaneous ferromagnetism is only stabilized if the density of states at the Fermi level is sufficiently high.  In these metallic regimes, the calculated N(E$_F$) 
remains too low, meaning the intra-atomic exchange interaction (I) fails to overcome the kinetic energy penalty of band shifting, thereby rendering the FM phase less 
stable than the unpolarized configuration.

In contrast, when the calculations are executed with the NUPDOWN = 0 constraint, the system achieves an entirely different electronic ground state. 
Rather than forming a conventional antiferromagnetic configuration with alternating antiparallel spins, individual Ru atoms consistently maintain a local spin 
magnetic moment of exactly 0 $\mu_B$ across all concentrations considered.  Physically, this indicates that the NUPDOWN = 0 specification quenches the 
exchange splitting entirely, forcing complete, localized electron pairing within the lower-energy crystal field levels.

This forces complete, localized electron pairing within the crystal-field split levels on every single ruthenium site, nominally satisfying an 
e$^4$($\uparrow\downarrow, \uparrow\downarrow$) t$_2^0$($\cdot \cdot \cdot$) electronic distribution. As displayed in the corresponding symmetric DOS graphs, 
this localized pairing leads to distinct electronic regimes as a function of concentration (Fig.\ref{fig12}).  For specific doping concentrations (x = 1 and 3), 
the system behaves as an insulator or semiconductor. The symmetric full occupancy of the lower-energy states clears out the immediate vicinity of the Fermi level 
(E$_F$ = 0 eV), leaving a well-defined energy gap.  Conversely, for x = 2 and the higher concentrations (x = 4, 5, 7, and 9), the system exhibits metallic character.   
Strong inter-dopant orbital hybridization causes the states to broaden and cross the Fermi level, creating a normal metallic phase with a finite, continuous density 
of states.  Notably, unlike the unstable FM phase, these states cross E$_F$ in a perfectly mirrored, spin-balanced manner, preventing any net exchange splitting 
or kinetic energy penalties.

By either shifting these electronic states away from Fermi level (as seen in the insulating regimes) or forcing them to cross it symmetrically (in the metallic 
regimes), this quenched non-magnetic configuration minimizes the total electronic energy of the crystal.  This structural organization perfectly explains why the 
NUPDOWN = 0 state maintains a significantly lower total energy compared to the FM phase across all concentrations.

Next, the electronic density of states of Rh$^{3+}$ (4d$^{6}$) doped w-AlN, shown in Figs.\ref{fig13} and \ref{fig14}, is analyzed.  As with the previous 
dopants, the host's tetrahedral crystal field splits the Rh$^{3+}$ 4d levels into lower-energy e and higher-energy t$_{2}$ sub bands.  However, the interplay 
between this local environment and the imposed spin constraints leads to a magnetic stability profile that differs from both the Cr- and Ru-doped systems.

Specifically, for the single dopant case (x = 1), total energy calculations reveal that the ferromagnetic (FM) state is favored over the non-magnetic (NUPDOWN = 0) state. 
As shown in the top-right panel of the AFM density of states in Fig.\ref{fig14}, forcing a zero local magnetic moment pins the Fermi level directly at the center of a 
sharp, symmetric Rh 4d impurity peak.  This density of states peak at the Fermi level induces electronic instability.

In contrast, the FM calculation (NUPDOWN = ND) yields a localized spin magnetic moment of approximately 1$\mu_B$ on the Rh$^{3+}$ ion, indicating a 
low-spin state.  Driven by the crystal field splitting, the six 4d electrons occupy the subshells unevenly as e$^4$($\uparrow\downarrow, \uparrow\downarrow$) 
t$_2^2$($\uparrow, \uparrow, \ \cdot$), where the lower-energy e doublet is closed and the remaining two electrons align parallel within the t$_{2}$ triplet.  
This spin polarization splits the impurity levels, shifting the core of the majority peak well below the Fermi level so that E$_{F}$ falls onto the 
shoulder of the majority spin-up band (Fig.\ref{fig13}).  By significantly reducing N(E$_F$) relative to the sharp AFM peak, this configuration 
minimizes the kinetic energy penalty via exchange splitting, thereby stabilizing the FM phase specifically at x = 1.

However, this energetic preference reverses completely for all higher concentrations (x = 2 to 9), where the NUPDOWN = 0 (AFM) configuration becomes the stable 
ground state.  In the NUPDOWN = ND (FM) calculation, increasing the Rh concentration broadens the impurity bands, forcing DOS peaks to cross the Fermi level, 
inducing instability.

Looking at the AFM DOS for x $\geq$ 2 (Fig.\ref{fig14}), it can be seen the system quenches the local magnetic moments to 0$\mu_B$ on all Rh atoms 
by forcing complete, localized electron pairing within the crystal field sub-levels, generating a closed-shell-like e$^4$($\uparrow\downarrow, \uparrow\downarrow$)
t$_2^2$($\uparrow\downarrow, \cdot, \cdot$) distribution. This localized orbital pairing triggers a favorable electronic redistribution that successfully clears out 
the volatile states at the Fermi level, opening a robust band gap (visible for x = 2, 4, 5) or a valley (x = 3, 7, 9).  This structural elimination of states at 
E$_{F}$ minimizes the overall electronic energy, rendering the non-magnetic AFM configuration relatively more stable for all x $\geq$ 2 concentrations.

\section{\label{con}CONCLUSION}
In this work, the magnetic properties of w-AlN doped with transition metals (M = Cr, Ru and Rh) were investigated using spin-polarized DFT calculations with 
Al$_{54-x}$M$_x$N$_{54}$ supercell models for $x$ = 1, 2, 3, 4, 5, 7, and 9.  Defect formation energies, calculated as a function of the Fermi level, predict that 
Cr$^{4+}$, Ru$^{4+}$, and Rh$^{3+}$ are the most stable charge states for Cr, Ru, and Rh atoms substituting at the Al site.  Based on these findings, subsequent 
calculations reveal that Cr$^{4+}$-doped AlN is more stable in the ferromagnetic (FM) state than in the antiferromagnetic (AFM) state.  Conversely, Ru$^{4+}$- and 
Rh$^{3+}$-doped AlN favor the AFM state over the FM spin state.  DOS analysis of the FM model for Cr-doped AlN indicates that the system behaves as a semiconductor 
up to a 5.56\% Cr concentration, with the Fermi level located in a local band gap.  As the Cr concentration increases, the material manifests as a high-spin half-metal 
between 7.41\% and 12.96\%, before transitioning into a normal metal at 16.67\%.  Conversely, the AFM model, which remains less favorable energetically than the FM 
model, exhibits a forced singlet at $x$=1 that violates Hund’s rule.  At $x$=2, it restores local high-spin states in an antiparallel configuration, yielding a perfectly 
symmetric DOS.  Both low concentrations experience a Stoner instability due to sharp DOS peaks at the Fermi level.  At intermediate concentrations ($x$=3 to 7), the 
system adopts a complex AFM state with unequal local moments and asymmetric DOS channels, before culminating in total spin quenching and a symmetric, non-magnetic 
metallic state at $x$=9 due to strong orbital overlaps.

Ru-doped w-AlN consistently favors a spin-quenched non-magnetic (NM) configuration, characterized by fully paired $(e^4_{\uparrow\downarrow} t^0_2)$ electrons that 
erase local magnetic moments, over the unstable high-spin ferromagnetic (FM) phase.  This spin-balanced state minimizes total energy, establishing an insulating gap 
at x=1, 3 and a stable, mirrored metallic phase at intermediate and higher concentrations, which avoids the severe kinetic energy penalties that destabilize the 
metallic FM phase.

Rh-doped w-AlN exhibits a unique concentration-dependent magnetic profile where the ferromagnetic (FM) phase is favored only at x=1, while the non-magnetic 
(NUPDOWN = 0) configuration stabilizes at all higher concentrations (x = 2 to 9).  At x=1, the system adopts a low-spin $e^4_{\uparrow\downarrow} t^2_{\uparrow}$ 
configuration, lowering the Fermi-level density of states to resolve a severe Stoner instability found in the zero-spin state.  Conversely, for all 
concentrations where x $\geq$ 2, strong orbital pairing completely quenches local magnetic moments into a closed-shell-like 
$e^4_{\uparrow\downarrow} t^2_{\uparrow\downarrow}$ state, clearing out volatile states at the Fermi level to create a stable band gap or valley that minimizes 
total electronic energy.

\end{document}